\input amssym.def
\documentstyle{article}
\title{Geometrical formulation for the Siegel superparticle}
\author{A.A. Deriglazov\thanks{E-mail: theordpt@fftgu.tomsk.su}\\
Department of Mathematical Physics, Politechnical Institute,\\
Tomsk, 634055, Russia\\
 and \\
A.V.Galajinsky\\
Department of Theoretical Physics, Tomsk State University,\\
Tomsk, 634050, Russia}
\date{}
\begin{document}
\maketitle

\abstract{In the superspace $z^M = (x^\mu,\theta_R,\theta_L)$ the
global symmetries for $d$ = 10 superparticle model with kinetic terms
both for Bose and Fermi variables are shown to form a superalgebra,
which includes the Poincar\'e superalgebra as a subalgebra. The
subalgebra is realized in the space of variables of the theory by a
nonstandard way. The local version of this model with off-shell
closed Lagrangian algebra of gauge symmetries and off-shell global
supersymmetry is presented. It is shown that the resulting model is
dynamically equivalent to the Siegel superparticle.}

\section{Introduction}

By now a problem of manifestly Poincar\'e covariant quantization of the
Green -- Schwarz superstring$^1$ and of massless relativistic
superparticles$^{2-5}$ has no fully satisfactory solution. In the
Lagrangian framework we are faced with the Siegel $k$-symmetry$^5$ of
the Lagrangians, which is written for spinor variables as
$\delta\theta_\alpha = \Pi_\mu (\Gamma^\mu k)_\alpha$. On the constraints
surface of the theories, only half of the $k$-parameters
contributes to $\delta\theta_\alpha$ by wirtue of the condition $\Pi^2 = 0$.
Since spinor $k_\alpha$ is chosen in the minimal spinor representation
of $SO(1,9)$ (we discuss $d$ = 10 case only), this leads in particular
to well-known difficulties in imposing a covariant gauge. Despite the
importance of $k$-symmetry, there is no its clear geometrical formulation
which would be suitable for quantization (in comparing with gauge or
general coordinate transformations).

In the Hamiltonian framework, there are eight first class constraints
corresponding to $k$-trans\-for\-mations, which are combined with eight second
class constraints in the following covariant equations ${\overline
p}_\theta + {\rm i}{\overline\theta}\Gamma^\mu p_\mu = 0$. To apply
standard methods of the Hamiltonian quantization, it is necessary to
divide the constraints in a fully covariant manner. This task was
solved by using covariant projectors for the cases of Green --
Schwarz superstring$^6$ and of massive superparticle$^4$ with
linearly dependent sets of first and second class constraints in the
results. Thus, a real problem in these cases is quantization of a
theory with infinite rank of reducibility$^{7-9}$.

One of possibilities to avoid the problems is to build modified and
more convenient for the quantization formulations $^{10-14}$ (in
particular, a necessity of some modifications follows from
absence of terms corresponding to the covariant propagators for Fermi
variables in these theories). Siegel superparticle$^3$ (SSP) is the least
radical modification of such a kind, where in comparing with the Brink --
Schwarz superparticle only first class constraints are retained. Let us
enumerate the relevant facts for $AB$-formulation (see Ref. 7 and
references therein). The defining system of constraints for SSP has the
form$^9$
\begin{eqnarray}
p_e &\approx& 0, \qquad \pi^2 \approx 0, \qquad {\overline
p}_\psi \approx 0, \qquad \pi_\mu ({\overline p}_R\Gamma^\mu
)_\alpha \approx 0;\\
{\overline p}_{L\alpha} &\approx& 0, \qquad {\overline
p}_{R\alpha} + {\rm i}\pi_\mu ({\overline\theta}_R\Gamma^\mu
)_\alpha - {\rm i}{\overline\theta}_{L\alpha} \approx 0;
\end{eqnarray}
where $(\pi_\mu,~{\overline p}_R,~{\overline p}_L,~p_e,
{}~{\overline p}_\psi)$ are canonically conjugate momenta for the
configuration space variables É$(x^\mu ,~\theta_{R\alpha},
{}~\theta_{L\beta},~e, ~\psi_{L\alpha})$ and $\Gamma^{11}\theta_{R,L} =
\pm\theta_{R,L}$. We use the Majorana representation for
$\Gamma^\mu$-matrices in $d$ = 10 and the notations from Ref. 15.
The first- and second-class constraints (Eqs. (1) and (2) respectively)
are separated in manifestly covariant manner and there are only 8
linearly-indepen\-dent among 16 constraints $\pi_\mu
({\overline p}_R\Gamma^\mu )_\alpha \approx 0$, as a consequence of
$\pi^2 \approx 0$. Thus, there are 16 dynamical variables in the
Fermi-sector of the model. Lagrangian action, reintroducing (1) and (2),
can be written in the form
\begin{eqnarray}
S &=& \int d \tau \left( \frac{1}{2e}\Pi_\mu \Pi^\mu + {\rm i}
{\dot{\overline\theta}}_R\theta_L\right),  \cr
{~}&{~}&{~} \\
\Pi^\mu &\equiv& {\dot x}^\mu - {\rm i}{\overline\theta}_R
\Gamma^\mu{\dot\theta}_R + {\rm i}{\overline\psi}_L \Gamma^\mu \theta_L,
\qquad {\dot x}^\mu \equiv \partial x^\mu /\partial\tau , \nonumber
\end{eqnarray}
and possesses a global supersymmetry in the standard realization
$\delta\theta_R = \epsilon_R$, $\delta x^\mu = {\rm i}
{\overline\epsilon}_R\Gamma^\mu\theta_R$ and local Siegel symmetry,
the latter being associated with the fermionic first-class constraints
from Eq. (1)
\begin{eqnarray}
\delta_k\theta_R &=& \frac{1}{e}\Pi_\mu\Gamma^\mu k_L, \qquad
\delta_k\theta_L = \frac{2}{e^2}k_L\Pi^2, \nonumber \\
\delta_kx^\mu &=& {\rm i}{\overline\theta}_L\Gamma^\mu k_L +
{\rm i}{\overline\theta}_R\Gamma^\mu (\delta_k\theta_R), \\
\delta_ke &=& -~\frac{4{\rm i}}{e}{\overline k}_L\Pi_\mu
\Gamma^\mu\psi_L, \qquad \delta_k\psi_L = {\dot k}_L.\nonumber
\end{eqnarray}
Thus, the SSP action contains the term corresponding to mixed
propagator for fermions and an additional Fermi variable $\psi_L$,
playing (as is seen from its transformation law) the role of a gauge
field for the $k$-symmetry.

On this ground, the question about studying an algebraic structure
of global transformations (the $\kappa$-algebra below) corresponding
to Siegel symmetry arises quite naturally. As will be shown, the latter
is the local and nonlinear version of the $\kappa$-algebra.

The paper is organized as follows. In Sec. 2 we analyze global
symmetries of the simplest $d$ = 10 action with kinetic terms both for
Bose and Fermi variables. We show that in the space of variables
of the theory it is realized a superalgebra which includes
$\kappa$-transformations and the  Poincar\'e superalgebra as subalgebras.
The last one is realized in a non-standard way. Note that, in
accordance with the Haag -- Lopuszanski -- Sohnius theorems$^{16}$
a superalgebra which is more wide than the super Poincar\'e one, can
be realized in nontrivial quantum theory by a special way only. Namely,
the generators of transformations added to the super Poincar\'e ones
have to vanish on physical states. This is exactly the case for
our model. In Sec. 3 we derive the local version for the
$\kappa$-transformations of the initial action which can be
done without lose of the global off-shell supersymmetry. The resulting
model is dynamically equivalent to SSP, but the structure of local
symmetries turns out to be more simple in comparing with (4) (in
particular, the Lagrangian algebra is off-shell closed). In Conclusion
we discuss possibilities of additional modifications with the aim
to get an equivalent to the Brink -- Schwarz superparticle formulation.

\section{${\Bbb R}^{10,32}$ superspace}

Consider $d =$ 10 superspace with the coordinates $z^M = (x^\mu,
\theta_R, \theta_L)$. Here the odd sector is parametrized by a
pair of Majorana -- Weyl spinors with opposite chirality:
$\Gamma^{11}\theta_{R,L} = \pm\theta_{R,L}$. Let us analyze global
symmetries which are present in the simplest Poincar\'e and
reparametrization invariant action in this superspace
\begin{equation}
S = \int d\tau\left( \frac{1}{2e}{\dot x}_\mu {\dot x}^\mu +
{\rm i}{\dot{\overline\theta}}_R \theta_L \right).
\end{equation}
Except the trivial translations $\delta_\epsilon\theta_L = {\rm i}
({\overline\epsilon}_L {\widetilde Q}_R)\theta_L \equiv \epsilon_L$
(where $\epsilon_L$ and ${\widetilde Q}_R$ are the global parameter and
the generator respectively), the model (5) is invariant under the
following transformations in the space of functions $z^M(\tau)$:
$\delta_\kappa z^M = {\rm i}({\overline\kappa}_LS_R)z^M$ where:
\begin{eqnarray}
\delta_\kappa\theta_L &=& 0, \qquad \delta_\kappa\theta_R =
\frac{1}{e}{\dot x}_\mu\Gamma^\mu\kappa_L,\cr
{~}&{~}& {~}\\
\delta_\kappa x^\mu &=& {\rm i}{\overline\theta}_L\Gamma^\mu
\kappa_L. \nonumber
\end{eqnarray}
A commutator of two $\kappa$-transformations yields a new one, which
can be written in the form
\begin{equation}
[\delta_{\kappa_1}, \delta_{\kappa_2}] z^M = \delta_bz^M =
-{\rm i}(b_\nu {\widetilde B}^\nu )z^M, \qquad b^\mu =
-{\rm i}{\overline\kappa}_1\Gamma^\mu\kappa_2,
\end{equation}
where
\begin{equation}
\delta_b\theta_R = \frac{1}{e}b_\mu\Gamma^\mu{\dot\theta}_L,
\qquad \delta_b\theta_L = \delta_bx^\mu = 0.
\end{equation}
The mixed commutators of the $\kappa$- and $\epsilon$-transformations
give a Poincar\'e translation $\delta_ax^\mu$ with the parameter $a^\mu
= {\rm i}{\overline\kappa}_L\Gamma^\mu\epsilon_L$. It is
straightforward to check that the commutators for $\delta_\epsilon$,
$\delta_\kappa$, $\delta_b$, $\delta_a$ form a closed algebra with the
Jacobi identities fulfilled.

Dimension $d = 10$ is unique in the sense that to get the Eqs. like (7)
it is necessary to use the Fiertz identities for Majorana -- Weyl
spinors (which are defined in $d =$ 2(Mod 8)). Thus, introducing
notations ${\widetilde P}_\mu$, $M_{\mu\nu}$ for the Poincar\'e generators
and ${\widetilde Q}_R$, $S_R$, ${\widetilde B}_\mu$ for generators of
the above written transformations, we can conclude that the symmetry
transformations of the action (5) realize some Lie superalgebra with
the following commutation relations (the standard Poincar\'e subalgebra
is omitted)
\begin{eqnarray}
&\{ S_{R\alpha},S_{R\beta}\} = -(\Gamma^\mu \Gamma^0)_{\alpha
\beta}{\widetilde B}_\mu , \quad \{ {\widetilde Q}_{R\alpha},
{\widetilde Q}_ {R\beta}\} = 0, \cr
&\{ {\widetilde Q}_{R\alpha}, S_{R\beta}\} = (\Gamma^\mu
\Gamma^0)_{\alpha\beta} {\widetilde P}_\mu, \quad
[M_{\mu\nu}, {\widetilde B}_\rho ] = {\rm i}(\eta_{\mu\rho}
{\widetilde B}_\nu - \eta_{\nu\rho}{\widetilde B}_\mu ),\\
&[M_{\mu\nu} , S_{R\gamma} ] = -~\frac{\mbox{i}}{\mbox{4}}
(\Gamma_{\mu\nu}S_R)_\gamma , \quad [M_{\mu\nu} ,
{\widetilde Q}_{R\gamma}] = -~\frac{\mbox{i}}{\mbox{4}}
(\Gamma_{\mu\nu}{\widetilde Q}_R)_\gamma . \nonumber
\end{eqnarray}
For further consideration it is useful to redefine the basis in the
algebra by the rules: ${\widetilde Q}_R \rightarrow Q_R = {\widetilde
Q}_R - S_R$, ${\widetilde P}_\mu \rightarrow P_\mu = {\widetilde P}_\mu
+ {\widetilde B}_\mu /2$, ${\widetilde B}_\mu \rightarrow B_\mu =
{\widetilde B}_\mu /2$. In this basis every element of the algebra has
the form ${\rm i}(\frac{1}{2}\omega^{\mu\nu}M_{\mu\nu} - a^\mu P_\mu +
{\overline\epsilon}_LQ_R + {\overline\kappa}_LS_R - b^\mu B_\mu )$,
with the following commutation relations for the generatorsÓ   Î  ÅÎÉ
\begin{eqnarray}
&&\left. \begin{array}{l}
\{Q_{R\alpha} ,Q_{R\beta}\} = -2(\Gamma^\mu\Gamma^0)_{\alpha
\beta}P_\mu , \quad [Q_\alpha ,P_\mu ] = [Q_\alpha ,B_\mu ] = 0, \cr
[M_{\mu\nu} ,Q_{R\alpha} ] = -~\frac{\mbox{i}}{\mbox{4}}(\Gamma_{\mu\nu}
Q_R)_\alpha ; \end{array} \right. \\
&&\left. \begin{array}{l}
\{S_{R\alpha} ,S_{R\beta}\} = -2(\Gamma^\mu\Gamma^0)_{\alpha
\beta}B_\mu , \quad [S_{R\alpha} ,P_\mu ] = [S_{R\alpha} ,B_\mu ] = 0, \cr
[M_{\mu\nu} ,S_{R\alpha} ] = -~\frac{\mbox{i}}{\mbox{4}}(\Gamma_{\mu\nu}
S_R)_\alpha ; \end{array} \right. \\
&&\{Q_{R\alpha},S_{R\beta}\} = (\Gamma^\mu\Gamma^0)_{\alpha
\beta} (P_\mu + B_\mu ).
\end{eqnarray}
The explicit realization of the superalgebra (10) -- (12) for the model (5)
is
\begin{eqnarray}
&&\left\{ \begin{array}{l}
\delta_\epsilon \theta_L = \epsilon_L, \\
\delta_\epsilon \theta_R = -~\frac{\mbox{1}}{\mbox{$e$}}{\dot x}_\mu
\Gamma^\mu\epsilon_L, \\
\delta_\epsilon x^\mu = {\rm i}{\overline\epsilon}_L
\Gamma^\mu\theta_L; \end{array} \right. \\
&&\delta_ax^\mu = a^\mu, \qquad \delta_a\theta_R = \frac{\mbox{1}}{\mbox{2$e$}}
a_\mu\Gamma^\mu{\dot\theta}_L, \qquad \delta_a\theta_L = 0; \\
&&\delta_\omega x^\mu = {\omega^\mu}_\nu x^\nu, \qquad
\delta_\omega\theta_{R,L} = -~\frac{1}{8} \omega_{\mu\nu}
\Gamma^{\mu\nu}\theta_{R,L}; \\
&&\left\{ \begin{array}{l}
\delta_\kappa\theta_L = 0, \\
\delta_\kappa\theta_R = \frac{\mbox{1}}{\mbox{$e$}}{\dot x}_\mu
\Gamma^\mu\kappa_L, \\
\delta_\kappa x^\mu = {\rm i}{\overline\theta}_L
\Gamma^\mu\kappa_L; \end{array} \right. \\
&&\delta_b\theta_R = \frac{1}{e}b_\mu\Gamma^\mu{\dot\theta}_L,
\qquad \delta_bx^\mu = \delta_b\theta_L = 0.
\end{eqnarray}

Let us give some comments concerning the structure of this superalgebra.

a) From an algebraic point of view, in the basis chosen the
superalgebra (10)--(12) contains two subalgebras ($M,P,Q$) and
($M,B,S$), both satisfying the commutation relations of the
Poincar\'e superalgebra$^{16}$, and having nontrivial overlapping
in the odd sector (12).

b) In Eqs. (13) -- (17) namely the ($M,P,Q$) subalgebra is
realized as the Poincar\'e superalgebra, because just the
$P^\mu$-generator corresponds to translations for $x^\mu$-variables.
Note that only on-shell $({\dot\theta}_L = 0)$ the Poincar\'e
translations (14) are realized in a standard way.

c) The transformations (17), corresponding to the $B_\mu$-generators,
vanish on-shell and consequently $\{S_{R\alpha},S_{R\beta}\}|_{\rm
on-shell} = 0$. This means that $S_{R\alpha}|a> = 0$ for any physical
state $|a>$ and, therefore, the ($S,B$)-subalgebra is a trivial symmetry
in quantum theory. However, such a construction turns out to be
useful for describing of superparticle models because the transformations
(16) for $\theta_R$ in fact are the linearized Siegel transformations
if we consider $\kappa_L$ as a local parameter.

Note also that the transformations (17) form a trivial symmetry
of the type $\delta\varphi^i = B^{[ij]}\delta S/\delta\varphi^j$ and
therefore are present in the action even with a local parameter
$b_\mu (\tau)$.

d) Since the action (5) is symmetric under the change $\theta_{R,L}
\rightarrow {\rm i}\theta_{L,R}$, it follows that there exists
conjugate to (10)--(12) superalgebra, which can be got from Eqs.
(10)--(12) by the change $L \leftrightarrow R$ (an explicit
realization of the conjugate algebra in the action (5) is achieved by
the change $\xi \rightarrow {\rm i}\xi$ for all odd parameters and
variables $\xi$). In the next section, the transformations (16) will
have been localized in (5) in such a way that the global supersymmetry
with a parameter $\epsilon_R$ from the conjugate superalgebra will be
present in the resulting action.

e) The superalgebra (10) -- (12) is realized by Eqs. (13) -- (17)
on the space of functions $z^M(\tau)$, because it is convenient for
our goals. Let us note that it may be realized on the superspace
$(x^\mu,\theta_L,\theta_R)$ as well, if one omits the multiplier 1/$e$
and $\tau$-derivatives in Eqs. (13) -- (17).

\section{The superparticle in ${\Bbb R}^{10,32}$-superspace}

{}From an analysis of constraints in the Hamiltonian formalism it
follows that there are 16 + 16 dynamical (phase) degrees of freedom in
the Fermi sector of the model (5), instead of 16 ones in the Siegel's
model. Therefore, let us consider a local version of the
transformations (16) in the action (5). The Noether procedure$^{17}$,
for example, can be used for these goals. In order to localize the
$\kappa$-transformations it is sufficient to covariantize the
derivatives ${\dot x}^\mu$ and to modify the transformation law for
$\delta_\kappa \theta_R$ by nonlinear, in the coordinates, terms. The
resulting locally-invariant action has the form
\begin{equation}
S = \int d\tau \left( \frac{1}{2e} {\rm D} x_\mu {\rm D} x^\mu
+ {\rm i}{\dot{\overline\theta}}_R \theta_L \right),
\end{equation}
where ${\rm D} x_\mu \equiv {\dot x}^\mu + {\rm i}{\overline\psi}_L
\Gamma^\mu \theta_L$ and $\psi_L$ is a Majorana -- Weyl spinor playing
the role of a gauge field. The local version for Eq. (16) looks like
$$
\delta_\kappa \theta_R = \frac{1}{e} {\rm D} x_\mu \Gamma^\mu
\kappa_L, \qquad \delta \psi_L ={\dot\kappa}_L,
$$
\makebox{~} \hfill (19)\\
$$
\delta_\kappa x^\mu = {\rm i}{\overline\theta}_L \Gamma^\mu
\kappa_L
$$
and forms off-shell closed algebra together with the local
$\delta_b$-transformations of Eq. (17).

In comparing with Eq. (5), the new terms in Eq. (18) don't allow to
achieve an invariance under the global supersymmetry transformations
(13) with the parameter $\epsilon_L$ or to obtain some generalization
of those formulae. However, analogous transformation for (13) in the
conjugate superalgebra (with a parameter $\epsilon_R$) is a symmetry of
the modified action (18). A mere realization of the symmetry has the
form
\stepcounter{equation}
$$
\delta_\epsilon \theta_R = \epsilon_R, \qquad \delta_\epsilon
\theta_L = \frac{1}{e} {\rm D} x_\mu \Gamma^\mu \epsilon_R,
$$
\makebox{~} \hfill (20)\\
$$
\delta_\epsilon x^\mu = {\rm i}{\overline\epsilon}_R \Gamma^\mu
\theta_R, \qquad \delta_\epsilon e = -2{\rm i}({\overline\psi}_L
\epsilon_R).
$$
So far as
\stepcounter{equation}
\begin{equation}
[\delta_{\epsilon_1}, \delta_{\epsilon_2}] \theta_L =
\frac{{\rm i}}{e} ({\overline\epsilon}_{1R} \Gamma_\mu \epsilon_{2R}
\Gamma^\mu ({\dot\theta}_R - \frac{1}{e} {\rm D} x_\nu
\Gamma^\nu \psi_L),
\end{equation}
the algebra of the commutators is closed on-shell only. Following the
standard ideology$^{17}$, it is not difficult to choose an auxiliary
variables, which provide off-shell closure of thus realized superalgebra
Poincar\'e. Let us introduce $d$ = 10 vector $h^\mu (\tau
)$ and modify the transformation law for $\theta_L$ in the following way:
$\delta_\epsilon \theta_L = \frac{1}{e} ({\rm D} x_\mu + h_\mu )
\Gamma^\mu \epsilon_R$. Requiring the off-shell closure for
commutators, one can find the transformation law for $h_\mu$. It is
straightforward to check that the transformations
\begin{eqnarray}
&&\delta_\epsilon \theta_R = \epsilon_R, \qquad \delta_\epsilon
\theta_L = \frac{1}{e} ({\rm D} x_\mu + h_\mu ) \Gamma^\mu
\epsilon_R, \nonumber \\
&&\delta_\epsilon x^\mu = {\rm i}{\overline\epsilon}_R \Gamma^\mu
\theta_R, \qquad \delta_\epsilon e = -2{\rm i}({\overline\psi}_L
\epsilon_R), \\
&&\delta_\epsilon h^\mu = {\rm i}{\overline\epsilon}_R \Gamma^\mu
\left( {\dot\theta}_R - \frac{1}{e} ({\rm D} x_\nu + h_\nu )
\Gamma^\nu \psi_L \right) \nonumber
\end{eqnarray}
form an off-shell realization of the superalgebra Poincar\'e together
with the Poincar\'e translations $\delta_ax^\mu = a^\mu$, $\delta_a
\theta_{L,R} = 0$. The variation of Eq. (18) under the (22) can be
written in the form
\begin{eqnarray}
\delta_\epsilon S &=& \int d\tau \left( -~\frac{\rm i}{e}
{\overline\epsilon}_R \Gamma^\mu ({\dot\theta}_R - \frac{1}{e}
{\rm D} x_\nu\Gamma^\nu\psi_L)h_\mu \right) \equiv \cr
&\equiv& \int d\tau \delta_\epsilon \left( \frac{1}{2e} h_\mu
h^\mu \right),
\end{eqnarray}
whence it follows a final expression for globally supersymmetric under
Eq. (22) and locally invariant under Eq. (19) action
\begin{equation}
S = \int d\tau \left( \frac{1}{2e} {\rm D} x_\mu {\rm D} x^\mu
+ {\rm i}{\dot{\overline\theta}}_R \theta_L - \frac{1}{2e} h_\mu
h^\mu \right).
\end{equation}
Note that after the redefinition $h^\mu \rightarrow eh^\mu$, the
variables $h^\mu$ are transformed as scalars under the
reparametrizations. Let us check that the obtained model is equivalent
to SSP. The constraints system, corresponding to the model (24) in the
Hamiltonian formalism is
\begin{eqnarray}
&&p_e \approx 0, \qquad \pi^2 \approx 0; \\
&&p_\psi \approx 0; \\
&&p_{h\mu} \approx 0, \qquad h_\mu \approx 0; \\
&&p_R \Gamma^\mu\pi_\mu \approx 0; \\
&&p_L \approx, \qquad p_R - {\rm i}{\overline\theta}_L \approx 0;
\end{eqnarray}
where ($\pi_\mu , p_R, p_L, p_\psi , p_e, p_h$) are momenta for the
variables ($x^\mu$, $\theta_R$, $\theta_L$, $\psi_L$, $e$),
respectively. The constraints (25) are standard and together with Eq.
(27) they mean that (in the light-cone gauge) there are the following
dynamical degrees of freedom in the Bose sector of the model: $x^-,
x^i, i = 1, \ldots, 8$. The first-class constraints (26) allow
us to impose the gauge $\psi_L \approx 0$, thus $\psi_L$ is a pure
gauge field. In the system of the first-class constraints (28), there
are only 8 linearly independent (as a consequence of $\pi^2 \approx
0$). Taking into account the second-class constraints (29), we can
conclude that there are 16 physical fermionic degrees of freedom in the theory.

Thus, in Eqs. (19), (22), and (24) we obtained the model dynamically
equivalent to SSP, but possessing more simple algebra of gauge
symmetries in comparing with (4). To achieve these, the crucial
observation is a nonstandard realization of the Poincar\'e superalgebra
in space of variables of the theory.

One can note that there is an another possibility
to achieve the off-shell global supersymmetry in the action (18). The
change ${\rm D}x_\mu \rightarrow \Pi_\mu \equiv {\dot x}^\mu -
{\rm i}{\overline\theta}_R \Gamma^\mu {\dot\theta}_R + {\rm i}\psi_L
\Gamma^\mu \theta_L$ leads to the SSP action (3), which is invariant
under supersymmetry transformations in the standard realization
$\delta\theta_R = \epsilon_R$, $\delta x^\mu = {\rm i}{\overline\epsilon}_R
\Gamma^\mu \theta_R$, and under the local transformations (4). As we
have shown above, this additional modification is not necessary.

\section{Conclusion}

In comparing with the Brink -- Schwarz superparticle, there are absent
eight second class constraints in the Siegel model. As a consequence,
it leads to undesirable negative norm states after quantization$^{10}$.
The action (3), therefore, can be considered only as a model one to study
a task of a manifestly covariant quantization of a theory with infinite
rank of reducibility. For this reason, in a series of papers an
intriguing possibility of modification of SSP, so as to get an
equivalent to the Brink -- Schwarz superparticle model, was considered
($ABC$ and $ABCD$-models$^{10-13}$). An analogous modification can be
done as well, if one starts from Eq. (24) instead of Eq. (3). In
particular, the formulation equivalent to the $ABC$-model arises after
adding the term Å$S_1 = \int d\tau {\overline\theta}_R \chi \theta_L$
to the action (24) (where $\chi_{\alpha\beta} = -\chi_{\beta\alpha}$ are
the Lagrange multipliers). The transformation laws for $\psi_L$ and
$\chi$ can be chosen in such a way that the full action maintains some
global supersymmetry. However, the additional first-class constraints
arising from the $S_1$-term (the same as in the $ABC$-model) are
quadratic in fermions and it leads to well-known difficulties both for
Dirac quantization and for BFV one$^{8,9}$.

Let us note, therefore, one more possibility. It is straightforward to check
that the addition of the term $S_2 = \int d\tau {\rm i}{\overline\lambda}_L
(\theta_R - x_\mu \Gamma^\mu\theta_L)$ to Eq. (24) leads to the theory
which is equivalent to the Brink -- Schwarz superparticle (technically,
the $S_2$-term supply an addition to the constraints system (25) --
(29) of 8 gauge conditions for Eq. (28) and of an additional 8
second-class constraints which are combined in one Lorentz-covariant
equation $\theta_R - x_\mu \Gamma^\mu \theta_L \approx 0$). Since there
are second-class constraints only, the problems with the
Lorentz-covariant gauge fixing is absent in this model. However,
covariance under the Poincar\'e translations takes place on the
equations of motion for dynamical variables only. Also, the
supertranslations are realized only on the $SO(8)$ component of
$\theta_L$ variables. We couldn't find additional modifications
leading to Poincar\'e invariant action.

Thus, in the present paper it has been shown that in constructing of $d$ =
10 superparticle models with terms corresponding to the covariant
propagator for Fermi variables, it is useful to consider the
superalgebras which are more wide than the superalgebra Poincar\'e. The
last one in this case is a subalgebra, and is realized in the space of
variables of the theory by a nonstandard way. In the case of the
Siegel's model it allows us to obtain a formulation with more simple
(in particular, off-shell closed) algebra of the Lagrangian gauge
symmetries.

\bigskip

\centerline{Acknowledgments}

\bigskip

The authors are sincerely grateful to S.M. Kuzenko for useful
discussions. This work is supported in part by ISF Grant No M2I000
and European Community Grant No INTAS-93-2058.

\centerline{References}

\bigskip

\noindent
1. M. Green and J. Schwarz, Phys. Lett. {\bf B136} (1984) 367.\\
2. L Brink and J. Schwarz, Phys. Lett. {\bf B100} (1981) 310.\\
3. W. Siegel, Class. Quant. Grav. {\bf 2} (1985) 95.\\
4. J.M. Evans, Nucl. Phys. {\bf B331} (1989) 711.\\
5. W. Siegel, Phys. Lett., {\bf B128} (1983) 397.\\
6. J.M. Evans, Phys. Lett. {\bf B233} (1989) 307.\\
7. I.A. Batalin, and E.S. Fradkin, Phys. Lett. {\bf B122} (1983) 157.\\
8. F. Ebler, M. Hatsuda, E. Laenen, W. Siegel, P. Yamron,
T. Kimura, and A. Mikovic, Nucl. Phys. {\bf B364} (1991) 67.\\
9. F. Ebler, M. Hatsuda, E. Laenen, W. Siegel, and P. Yamron,
Phys. Lett. {\bf B254} (1991) 411.\\
10. A. Mikovic and W. Siegel, Phys. Lett. {\bf B209} (1988) 47.\\
11. W. Siegel, Phys. Lett. {\bf B203} (1987) 79.\\
12. M.B. Green, C. Hull, Mod. Phys. Lett. {\bf A5} (1990) 1399.\\
13. I. Bengtsson, Phys. Rev. {\bf D39} (1989) 1158.\\
14. A.A. Deriglazov, J. Nucl. Phys. {\bf 55} (1992) 3361; Int.
J. Mod. Phys. {\bf A8} (1993) 1093.\\
15. L. Brink and M. Henneaux, in {\it Principles of String Theory}.
New York /London: Plenum Press, 1988.\\
16. R. Hagg, J. Lopuszanski, and M. Sohnius, Nucl. Phys. {\bf B88} (1975)
61.\\
17. P. van Nienwenhuizen, Phys. Rep. {\bf 68} (1981) 189.

\end{document}